\documentclass[prb,twocolumn,showpacs,superscriptaddress]{revtex4}
\usepackage[latin1]{inputenc}
\usepackage{graphicx}
\usepackage{amssymb}
\usepackage{amsmath}
\usepackage{xspace}
\usepackage{dcolumn}
\usepackage{bm}
\usepackage{color}
\usepackage[extra]{tipa}

\newfont{\tensy}{cmsy10}

\newcommand{\ie}[0]{i.e.\@\xspace}
\newcommand{\eg}[0]{e.g.\@\xspace}
\newcommand{\etal}[0]{et al.\@\xspace}

\newcommand{\oP}{\hat{P}}
\newcommand{\oQ}{\hat{Q}}

\newcommand{\on}{\hat{n}}

\newcommand{\oh}{\mbox{$\frac{1}{2}$}}

\newcommand{\om}[0]{\omega}
\newcommand{\Ep}{\varepsilon_\text{p}}

\newcommand{\kF}{k_\text{F}}
\newcommand{\kB}{k_\text{B}}
\newcommand{\nag}{{\phantom{\dag}}}

\newcommand{\las}[0]{\langle}
\newcommand{\ras}[0]{\rangle}

\newcommand{\la}[0]{\left\las}
\newcommand{\ra}[0]{\right\ras}
\newcommand{\ket}[1]{\left|#1\ra}
\newcommand{\bra}[1]{\la#1\right|}

\newcommand{\en}[0]{\epsilon}

\begin{document}


\title{%
Dynamic charge correlations near the Peierls transition}

\author{Martin Hohenadler}
\affiliation{%
Institut f\"ur Theoretische Physik und Astrophysik, Universit\"at W\"urzburg,
97074 W\"urzburg, GER}

\author{Holger Fehske}
\affiliation{%
Institut f\"ur Physik, Ernst-Moritz-Arndt-Universit\"at Greifswald, 17489
Greifswald, GER}

\author{Fakher F. Assaad}
\affiliation{%
Institut f\"ur Theoretische Physik und Astrophysik, Universit\"at W\"urzburg,
97074 W\"urzburg, GER}

\begin{abstract}
  The quantum phase transition between a repulsive Luttinger liquid and an
  insulating Peierls state is studied in the framework of the one-dimensional
  spinless Holstein model. We focus on the adiabatic regime but include the full
  quantum dynamics of the phonons. Using continuous-time quantum Monte
  Carlo simulations, we track in particular the dynamic charge structure factor and
  the single-particle spectrum across the
  transition. With increasing electron-phonon coupling, the dynamic charge structure
  factor reveals the emergence of a charge gap, and a clear signature of
  phonon softening at the zone boundary. The single-particle spectral
  function evolves continuously across the transition. Hybridization of the 
  charge and phonon modes of the Luttinger liquid description leads to two
  modes, one of which corresponds to the coherent polaron band. This band
  acquires a gap upon entering the Peierls phase, whereas the other mode
  constitutes the incoherent, high-energy spectrum with backfolded shadow
  bands. Coherent polaronic motion is a direct consequence of quantum lattice
  fluctuations. In the  strong-coupling regime, the spectrum is described by the static,
  mean-field limit. Importantly, whereas finite electron density in general
  leads to screening of polaron effects, the latter reappear at half filling
  due to charge ordering and lattice dimerization.
\end{abstract} 

\date{\today}

 
\maketitle

\section{Introduction}\label{sec:intro}

The Peierls transition is a hallmark feature of quasi-one-dimensional
materials, such as conjugated polymers or organic charge-transfer salts. 
Peierls originally predicted a band insulator for any finite electron-phonon
coupling for a half-filled system due to an instability in the presence of
perfect nesting.\cite{Peierls} A full understanding of the physics requires
to account for both quantum lattice fluctuations and strong electronic
correlations.\cite{TsNaYaSi90} For example, it has been found theoretically
that quantum fluctuations of the lattice give rise to a metallic state below
a finite critical coupling strength.\cite{HiFr82} 

A minimal yet rich theoretical setting studied is the spinless
Holstein model\cite{Ho59a} at half filling. With increasing
electron-phonon coupling, a quantum phase transition from a metal to a Peierls
insulator (PI) with charge-density-wave order occurs.\cite{HiFr82,HiFr83II}
The phase diagram has been mapped out with high a accuracy.\cite{BuMKHa98,WeFe98,JeZhWh99,HoWeBiAlFe06} The character of the
phase transition depends on the phonon frequency.\cite{HoWeBiAlFe06} In the
adiabatic regime considered here, the Peierls state is a band
  insulator, and the transition is accompanied by the softening of the
  phonon mode at the zone boundary. The critical coupling is rather small,
so that our Monte Carlo method becomes very efficient (see Sec.~\ref{sec:method}).
In the nonadiabatic
regime, small polarons form a polaronic superlattice, and the bare phonon
mode hardens on approaching the transition.\cite{HoWeBiAlFe06}

A metallic phase of one-dimensional spinless fermions is described by Luttinger
liquid (LL) theory, with an interaction parameter $K_\rho$, even in the
presence of a coupling to phonons.\cite{0022-3719-14-19-010,MeScGu94}
Until recently, from scaling analysis of the ground-state
energy,\cite{MKHaMu96,FeHoWe00,WeFeWeBi00} there was evidence for a crossover
from an attractive LL ($K_\rho>1$) at low phonon frequencies, to a repulsive
LL ($K_\rho<1$) at high phonon frequencies. However, large-scale
density-matrix renormalization group (DMRG)
calculations of the charge structure factor have revealed that the LL is always repulsive, \ie
$K_\rho\leq1$.\cite{Ej.Fe.09} The Peierls quantum phase transition
in the spinless Holstein model is of the Kosterlitz-Thouless (KT) type for
all values of the phonon frequency, with $K_\rho=1/2$ at the transition
point.\cite{Ej.Fe.09} The KT transition was previously known to occur in the nonadiabatic regime
where the model can be mapped onto the $XXZ$ Hamiltonian.\cite{HiFr82,HiFr83II,BuMKHa98} The suppression of
superconducting correlations is ascribed to the spinless nature of the charge
carriers, which prevents onsite pairing. In the
spinful model at quarter filling, a transition from a LL to a (metallic)
Luther-Emery liquid with finite spin gap but zero charge gap was
discovered.\cite{assaad:155124} An LL theory for gapless
Peierls states has been developed.\cite{Voi1998}

Given the continuous nature of the Peierls transition, one may argue that LL
features should also survive at moderate couplings in the PI state, \ie
sufficiently close to the transition. Such a scenario will indeed emerge from
our results for the single-particle spectral function, which reveal mixed charge and
phonon modes (\ie, polarons) in both phases, with a finite single-particle
gap in the Peierls phase. For the metallic phase, we shall compare to exact
bosonization results for the spectral function.\cite{MeScGu94}

Dynamic properties of the spinless Holstein model have been studied
numerically and
analytically,\cite{HoAivdL03,SyHuBeWeFe04,HoNevdLWeLoFe04,HoWeAlFe05,CrSaCa05,WeBiHoScFe05,HoWeBiAlFe06,SyHueBe06,LoHoFe06,LoHoAlFe07}
(we refer to these papers for a more comprehensive review of work
on the spinless Holstein model) except for the dynamic charge structure
factor. The spinless Holstein model may be considered as the strong-Hubbard-$U$ limit
of the Hubbard-Holstein model, so that the formation of singlet bipolarons is suppressed.
Away from half filling, it captures polaronic effects believed
to be crucial in strongly correlated materials such as one-dimensional MX
chains,\cite{BB87,BB88,BS93} or  quasi-two respectively three-dimensional
 manganites.\cite{David_AiP,HoEd01,HoNevdLWeLoFe04}

Here we apply an exact continuous-time quantum Monte Carlo (CTQMC) method to revisit
the Peierls transition in the spinless Holstein model. Apart from presenting
a highly non-trivial test of the method for lattice problems, we make several
contributions to the understanding of the physics near the transition.  In
particular, we compute the dynamic charge structure factor, and relate the
single-particle spectral function to LL theory and polaron physics on both
sides of the transition, and to mean-field theory in the strong-coupling regime.

The paper is organized as follows. We introduce the model in
Sec.~\ref{sec:model}, and provide some details of the method for
the present problem in Sec.~\ref{sec:method}. A discussion of our results is
given in Sec.~\ref{sec:results}, and we conclude in Sec.~\ref{sec:conclusions}.

\section{Model}\label{sec:model}

The Hamiltonian $\hat{H} = \hat{H}_0 + \hat{H}_1$ takes the form
\begin{align}\label{eq:model}
  \hat{H}_0 &= 
  -t \sum_{\las ij\ras} \left(c^\dag_{i}c^\nag_{j} + \text{H.c.}\right)
  -\mu \sum_i \on_i\,,
  \\\nonumber
  \hat{H}_1 &= 
  \sum_i \left(\frac{1}{2M}\oP_i^2 + \frac{K}{2}\oQ_i^2\right)
  - \gamma \sum_i \oQ_i (\on_i-\oh)\,.
\end{align}
The first term $\hat{H}_0$ describes hopping of spinless fermions
between neighboring sites with overlap integral $t$ and free dispersion
$\epsilon(q)=-2 t\cos q$, $\mu$ is the chemical
potential, and the charge density operator is defined as $\on_i=c^\dag_i c^\nag_i$ with eigenvalues 0, 1. The first part of $\hat{H}_1$
describes the free dynamics of the lattice in the harmonic
approximation; $\omega_0$ is the phonon frequency, and $K=\omega_0^2 M$. 
The second part couples the electron density to the lattice displacement.
The half filled band (density $n=0.5$) corresponds to $\mu=0$, and we work in
one dimension.

Due to the use of first quantization (more convenient for the present method), the
notation, in particular the coupling $\gamma$, differs from that used
in previous work we shall compare to,\cite{HoWeBiAlFe06,SyHuBeWeFe04} where
the coupling term is written in the form $-g\om_0 \sum_i (b^\dag_i +
b^\nag_i)(\on_i-\oh)$. To reconcile the two notations, we use the dimensionless coupling
constant $\lambda= 2\Ep/W$, where $\Ep$ is the polaron binding energy, and
$W=4t$ is the bandwidth; $\lambda$ is the relevant dimensionless ratio in the
adiabatic regime $\om_0< t$, separating weak ($\lambda\ll1$) and strong
coupling ($\lambda\gg1$); small-polaron formation occurs at
$\lambda\simeq1$.\cite{dRLa82} In terms of $g$ we have $\Ep=g^2\om_0$ and 
$\lambda=g^2\om_0/(2t)$, whereas in terms of $\gamma$ the relations are 
$\Ep=\gamma^2/(2K)=\gamma^2/(2\om_0^2)$ and $\lambda=\gamma^2/(\om_0^2 W)$.
We use $t$ as the unit of energy, and set $\hbar=\kB=M=1$.

\section{Method}\label{sec:method}

The CTQMC is based on
an exact diagrammatic
expansion of the partition function around the non-interacting
limit;\cite{Ru.Sa.Li.05} the expansion converges for any finite fermionic
system at finite temperature. 
The fact that Wick's theorem holds for each
configuration allows for a simple calculation of one and two-particle fermionic
Green's functions. As discussed elsewhere,\cite{Ru.Sa.Li.05,assaad:035116}
updates take the form of the addition and
removal of single vertices, and optionally flipping Ising spins. 
The method is exact also for strong coupling,
but becomes numerically less efficient due to large matrix sizes.
The numerical effort scales with the cube of the average expansion order; the
latter depends linearly on the system size  $N$, inverse temperature $\beta=1/T$ and
(effective) coupling strength.

The extension of the CTQMC method to electron-phonon problems
has been discussed before.\cite{assaad:035116}  Most importantly, as the
algorithm is action based,\cite{Ru.Sa.Li.05} the path integral over the
phonon degrees of freedom can be performed exactly for a wide range of
problems\cite{Feynman55} (see also Ref.~\onlinecite{assaad:035116}). This leads to a retarded
electron-electron interaction of the form
\begin{equation}
S_1=-\int_0^\beta \hspace*{-0.5em} \int_0^\beta 
d\tau d\tau'
\sum_i [n_i(\tau)-\oh] D(\tau-\tau')[n_i(\tau')-\oh]\,,
\end{equation}
where $D(\tau)$ is the phonon propagator of the Holstein model [diagonal in
real-space due to the onsite interaction in Eq.~(\ref{eq:model})], and
$n_i(\tau)$ is a Grassmann bilinear. The result
is a purely fermionic algorithm, with no additional sampling of phonon degrees of
freedom. The vertices have equal real space coordinates but
in general different times $\tau,\tau'$, distributed $\sim D(\tau-\tau')$. The range of
the retarded interaction along the imaginary time is of the order $1/\om_0$,
but we find the algorithm  to be efficient even for small $\om_0$.
The weight to add a vertex is proportional to $\lambda\beta N$.
In the present case, the spinless nature of the model makes only one spin
direction appear in the algorithm. There is no sign problem, and we consider
system sizes $N=4l+2$ with periodic boundary conditions.

Compared to previous QMC studies of the spinless Holstein
model,\cite{HiFr82,HiFr83II,KeHaMu96,HoNevdLWeLoFe04,CrSaCa05} the
CTQMC algorithm is free of errors associated with a Trotter discretization.
There is also no cutoff for the phonon Hilbert space, as the lattice degrees
of freedom are integrated out exactly.

Two key quantities for studying the Peierls transition are the static charge
structure factor
\begin{equation}
S_\rho(q) = \sum_{r} \left(\las \on_r \on_0 \ras - \las \on_r\ras \las\on_0\ras \right)  e^{iq r}\,,
\end{equation}
and the dynamic charge structure factor given in the Lehmann representation
as
\begin{align}\nonumber
  S_\rho(q,\om)
  =
  \frac{1}{Z}\sum_{mn} &{|\bra{n} \hat{\rho}_q \ket{m}|}^2
  e^{-\beta E_m} 
  \\
  &\times\delta(E_m-E_n-\om)
  \,,
\end{align}
where $\ket{n}$ is an eigenstate with energy $E_n$, and we have defined 
$\hat{\rho}_q = \sum_r e^{iqr} (\on_{r} - \las \on_r\ras) /\sqrt{N}$.

We also consider the single-particle spectral function,
\begin{align}\nonumber\label{eq:akw}
  A(q,\om)
  =
  \frac{1}{Z}\sum_{mn}
  &{|\bra{n} c_q \ket{m}|}^2 (e^{-\beta E_n}+e^{-\beta E_m})\\
  &\times\delta(E_m-E_n-\om)
  \,.
\end{align}
For the continuation from imaginary time to real frequencies, we use the
stochastic Maximum Entropy method.\cite{Be.04}

\section{Results}\label{sec:results}

\begin{figure}
  \includegraphics[width=0.45\textwidth,clip]{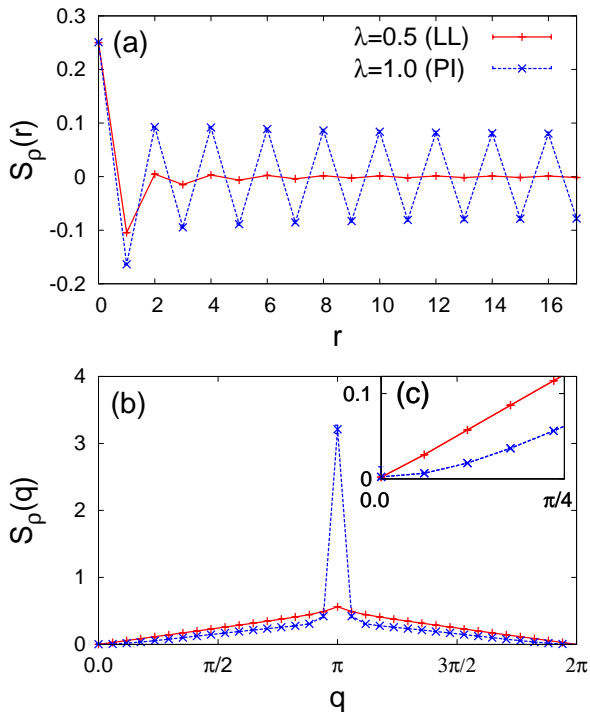}
  \caption{\label{fig:static} (Color online) Density-density correlations 
    $S_\rho(r)=\las\on_r\on_0\ras$ (a) and charge structure factor
    (b). The coupling $\lambda=0.5$ corresponds to the
    {\it metallic Luttinger liquid (LL) phase}, whereas $\lambda=1.0$ is in the
    {\it insulating Peierls phase (PI)}.  (c) Closeup of (b). Here $\beta t=N=34$, $\om_0=0.4t$ and
    $n=0.5$ (half filling). Lines are guides for the eye.}
\end{figure}

We anchor our choice of parameters to previous work,\cite{SyHuBeWeFe04,HoWeBiAlFe06} focusing on the
adiabatic regime $\om_0<\ t$ (typically realized in experiments)
where the Peierls transition falls into the weak or intermediate coupling
regime. In the latter, the CTQMC is by its nature particularly efficient due
to low average expansion orders.\footnote{The critical coupling $\lambda_\text{c}$ for the Peierls transition
scales with $\om_0$; see \eg Fig.~1 in
Ref.~\onlinecite{HoWeBiAlFe06}. Besides, charge carrier renormalization
(dressing) is much more pronounced for high phonon frequencies.}
Explicitly, following Ref.~\onlinecite{HoWeBiAlFe06}, we take $\om_0=0.4t$. We mainly
consider two characteristic values of the coupling strength, namely $\lambda=0.5$
and 1 (equivalent to $g^2=2.5$ respectively $5$). The latter lie just above
and below the Peierls transition, see Fig.~1 in
Ref.~\onlinecite{HoWeBiAlFe06}, with the critical coupling known from
DMRG calculations\cite{HoWeBiAlFe06} as $\lambda_\text{c}\approx0.7$ ($g_\text{c}^2\approx 3.5$).
The same parameters have been considered in
Refs.~\onlinecite{HoNevdLWeLoFe04,WeBiHoScFe05,SyHueBe06}. We shall also
compare to the case $\om_0=0.1t$ considered in
Ref.~\onlinecite{SyHuBeWeFe04}, for which $\lambda_\text{c}\approx 0.4$. The inverse temperature is set to
$\beta t=N$.

\subsection{Static charge correlations}

In Fig.~\ref{fig:static}(a) we show the density-density correlator
$S_\rho(r)=\las\on_r\on_0\ras$. The $T=0$ bosonization result is\cite{Giamarchi_book_04}
\begin{equation}\label{eq:corr}
  \las \on_x \on_0 \ras 
  =
  -
  \frac{K_\rho}{2\pi^2 x^2}
  +  
  \frac{A}{x^{2 K_\rho}}   \cos(2\kF x)\,,
\end{equation}
with $\kF=n\pi$ for the spinless case, equal to $\pi/2$ for half filling. In
the metallic phase, the LL parameter can be extracted from
\begin{equation}\label{eq:krhofit}
  K_\rho
  =
  2\pi \lim_{q\to 0} S_\rho(q)/q\,.
\end{equation}
For simplicity, we take $K_\rho\approx 2\pi S_\rho(q_1)/q_1$, where $q_1$ is
the smallest nonzero wavevector for a given lattice size.

In the metallic LL phase, the amplitude of density correlations shows a
power-law decay in real space.  Since $K_\rho=0.96(1)<1$ for $\lambda=0.5$,
we see a cusp in $S_\rho(q)$ at $q=2\kF$. The linear form of $S_\rho(q)$ at small
$q$ [Fig.~\ref{fig:static}(c)] is evidence for a metallic state at
$\lambda=0.5$. 

In the Peierls phase ($\lambda=1$), the
density correlation functions display long-range order (a charge
density wave with one fermion on every other site, and a staggered lattice
displacement with periodicity $\pi$). The small-$q$ behavior of $S_\rho(q)$ becomes nonlinear,
see Fig.~\ref{fig:static}(c). Via the continuity equation, the long-wavelength
limit of $S_\rho(q)$ can be directly related to the Drude part of the
optical conductivity,\cite{assaad:155124} and the absence of linear behavior of $S_\rho(q)$ as
$q\to0$ is equivalent to a vanishing Drude weight in the insulating Peierls
state. As a result of long-range charge-density-wave order, $S_\rho(q)$
diverges with system size at $q=\pi=2\kF$, and serves as an order
parameter. The Peierls state is quickly suppressed away from the commensurate
density $n=0.5$.\cite{HoHaWeFe06}

\begin{figure}
 \includegraphics[width=0.45\textwidth,clip]{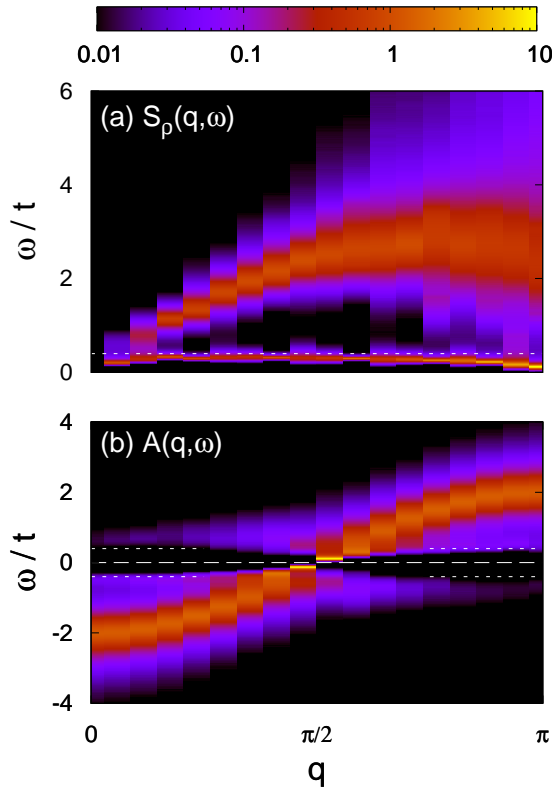}
 \caption{\label{fig:dynamics_LL} (Color online) (a) Dynamic charge structure
   factor $S_\rho(q,\om)$ and (b) single-particle spectral function
   $A(q,\om)$ in the {\it metallic LL phase}. Horizontal lines indicate (a)
   the bare phonon frequency $\om_0$ and (b) the chemical potential $\mu=0$
   (long dash) and $\pm\om_0$ (short dash). We have limited the intensity
   range to reveal the subdominant features; the full range for (a) is shown
   in Fig.~\ref{fig:dynamics_softening}(b); for (b) see also
   Fig.~\ref{fig:dynamics_LL2}. Parameters are $\om_0=0.4t$,
   $\lambda=0.5$, and $\beta t=N=34$.}
\end{figure}

The LL parameter $K_\rho$ has been calculated as a function of $g$ with high
accuracy in a large-scale DMRG study including finite-size scaling with up to 256
sites.\cite{Ej.Fe.09} As the CTQMC is not capable of studying such large systems, and due to
the additional complication of finite temperatures in our method
[Eqs.~(\ref{eq:corr}) and (\ref{eq:krhofit}) only hold at $T=0$] we do not
attempt a quantitative comparison here. Instead, we have checked for both
$\om_0=1$ and $0.1$ that even on small systems ($N=22$) the CTQMC correctly
reproduces $K_\rho\leq1$ (repulsive LL), albeit with significant finite-size
effects near $g_\text{c}$ as expected for a KT phase transition.\cite{Ej.Fe.09}

\subsection{Excitation spectra in the metallic phase}

We now turn to the discussion of the dynamical correlation functions, namely
the dynamic charge structure factor and the single-particle spectral
function.  To our knowledge, the former has not been calculated for the
present model before.  In all intensity plots, we have refrained from using
any kind of interpolation. This makes the limited (yet significantly higher
than in previous numerical work) momentum resolution  well
discernible.  The energy resolution was taken to be $0.01t$ or better.

The spectra in the LL phase ($\lambda=0.5$) are shown in
Fig.~\ref{fig:dynamics_LL}. The dynamic charge structure factor $S_\rho(q,\om)$
[Fig.~\ref{fig:dynamics_LL}(a)] shows the familiar continuum of particle-hole
excitations of fermions in one dimension. At long wavelengths, we
have a linear, gapless mode with dispersion $N(q,\om)\approx
N(q)\delta(\om-v_\rho q)$ expected for the metallic phase. Near the zone boundary, the
width of the continuum approaches the bare bandwidth $W=4t$.
In addition,
because of the coupling between the electron density and the phonons in the
Hamiltonian~(\ref{eq:model}), a signature of the renormalized optical phonon
mode is visible at low energies. Already below the Peierls transition at
$\lambda_\text{c}\approx0.7$, it shows partial softening near $q=\pi$.
Our findings are in excellent agreement with previous calculations
of the phonon spectral function,\cite{HoWeBiAlFe06,SyHuBe05}  see also 
Figs.~\ref{fig:dynamics_softening}(a) and (b) discussed below. Hence, despite the
nondispersive nature of the Einstein phonons in Eq.~(\ref{eq:model}) the
coupling gives rise to a momentum dependence of the renormalized phonon frequency.

\begin{figure}
 \includegraphics[width=0.45\textwidth,clip]{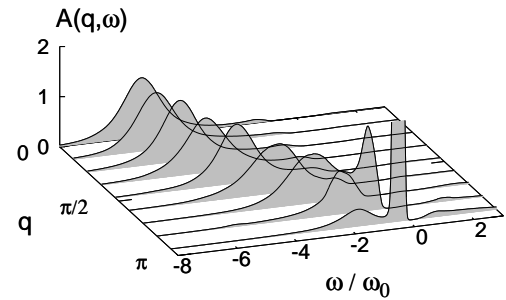}
  \caption{\label{fig:dynamics_LL2} (Color online) Single-particle spectral
    function $A(q,\om)$ in the {\it metallic LL phase}. Parameters as in
    Fig.~\ref{fig:dynamics_LL}(b).}
\end{figure}

The single-particle spectrum $A(q,\om)$ in the LL phase
[Fig.~\ref{fig:dynamics_LL}(b)] reveals a dominant band (note the logarithmic
scale in all our intensity plots) closely tracking
the cosine dispersion of the noninteracting problem, $\en(q)$.
Close to the Fermi level, inside the coherent interval $\om\in[-\om_0,\om_0]$, the main peak
becomes noticeably sharper.\cite{LoHoFe06} As noted
previously,\cite{MeScGu94,HoWeBiAlFe06,SyHuBeWeFe04} the incoherent part
($|\om|\geq\om_0$) of the spectrum is broadened due to multiphonon
processes;\cite{LoHoFe06} the width is roughly $g^2=2.5$. 

As expected, but not discussed in the numerical
literature\cite{SyHuBeWeFe04,HoWeBiAlFe06} (where the spectra are
mostly interpreted in terms of polaron theory), the spectrum shares the
main features of the exact bosonization result for a linear fermion
dispersion.\cite{MeScGu94} For comparison, we show the results in
Fig.~\ref{fig:dynamics_LL}(b) again in Fig.~\ref{fig:dynamics_LL2} 
using a representation similar to that in Ref.~\onlinecite{MeScGu94}. The bare charge and phonon modes of the LL
Hamiltonian of the spinless model hybridize and broaden; see also
Ref.~\onlinecite{assaad:155124}. This is equivalent to the formation of coherent
polaronic quasiparticles, \ie electrons dressed with phonons, with increased
effective mass. At $\kF=\pi/2$ (we focus on $\om<0$), there is a
dominant zero-energy peak and a phonon satellite with threshold energy
$-\om_0$ whose width depends on the coupling strength. With $q$ moving away from $\kF$, spectral weight
is transferred to the ``phonon peak'' (in the nomenclature of Meden
  \etal\cite{MeScGu94}) which eventually becomes the dominant
excitation and follows $\en(q)$. Since in contrast to
Ref.~\onlinecite{MeScGu94} the bandwidth $W\gg\om_0$, we observe multiple phonon
satellites which merge into a broad band. This picture is consistent with previous
numerical\cite{SyHuBeWeFe04,HoWeBiAlFe06} and analytical results\cite{SyHueBe06}
(though difficult to see in the latter) in the
metallic phase. Similar results and an explicit comparison to LL theory
have been reported for the spinful case.\cite{assaad:155124}

\begin{figure}
 \includegraphics[width=0.45\textwidth,clip]{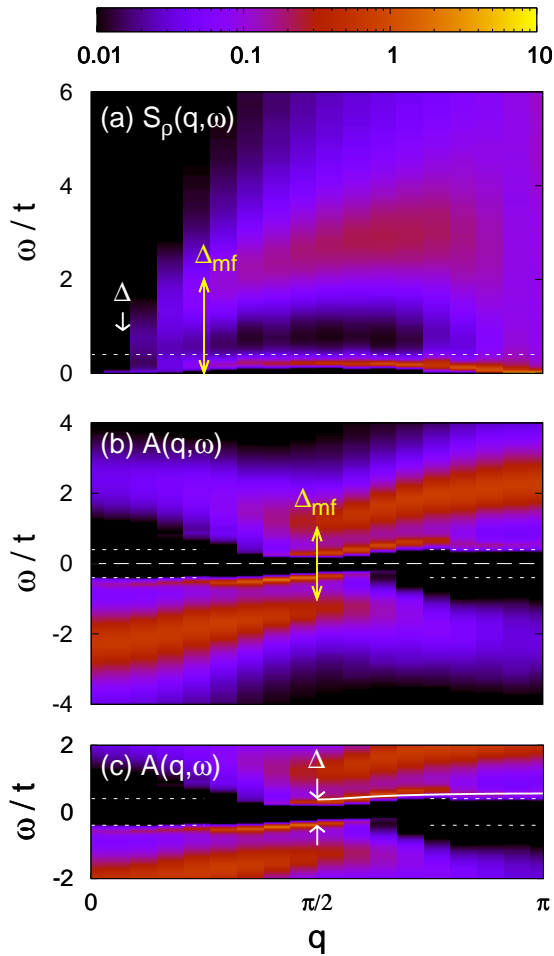}
  \caption{\label{fig:dynamics_Peierls} (Color online) (a) Dynamic charge structure
    factor $S_\rho(q,\om)$ and (b) single-particle spectral function $A(q,\om)$
    in the {\it Peierls phase}. (c) Closeup of (b). Horizontal lines indicate (a) the
    bare phonon frequency $\om_0$, (b) the chemical potential $\mu=0$ (long
    dashed) and  $\pm\om_0$ (short dashed), and (c)  $\pm\om_0$. The solid
    line in (c) is the shifted single-polaron band dispersion 
    $\epsilon_\text{p}(q)+\Delta/2$ for the same parameters,\cite{LoHoAlFe06}
    mapped from $[0,\pi]$ to $[\kF,\pi]$ to facilitate comparison with half filling.
    The double-headed arrows in (a) and (b)
    designate the mean-field gap $\Delta_\text{mf}=\Ep=2t$. The single-headed arrows
    in (a) and (c) indicate the Peierls gap $\Delta\approx 0.75t$. In (a) we
    have limited the intensity range to reveal the subdominant
    features; the full
    range is shown in Fig.~\ref{fig:dynamics_softening}(c). Parameters are
    $\om_0=0.4t$, $\lambda=1.0$, and $\beta t=N=34$.}
\end{figure}

\begin{figure}
 \includegraphics[width=0.45\textwidth,clip]{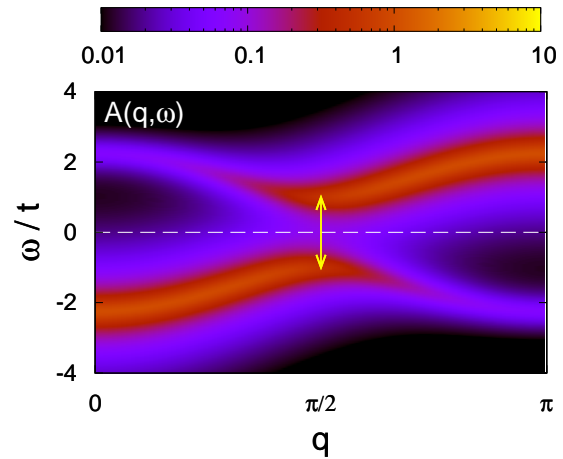}
  \caption{\label{fig:dynamics_Peierls_static} (Color online) Mean-field
    result for the single-particle spectrum  $A(q,\om)$ of the
    adiabatic ($\om_0=0$), spinless Holstein model at $T=0$. Here 
    $\Delta_\text{mf}=2t$ (arrow); we have replaced the $\delta$ peaks by
    Lorentzians with scaling parameter $0.3t$.}
\end{figure}

\subsection{Excitation spectra in the Peierls phase}

In Fig.~\ref{fig:dynamics_Peierls} we report the dynamic charge structure factor
and the single-particle spectrum in the Peierls state, at
$\lambda=1$. Mean-field results for $A(q,\om)$ to be compared to
Fig.~\ref{fig:dynamics_Peierls}(b) are shown in Fig.~\ref{fig:dynamics_Peierls_static}.

In the case of $S_\rho(q,\om)$, see Fig.~\ref{fig:dynamics_Peierls}(a), we note that
almost the entire spectral weight is located close to $\om=0$, $q=\pi$.  All
other features, including the particle-hole continuum, have a spectral weight
that is at least an order of magnitude lower (see caption). The accumulation of spectral weight at
$\om=0$, $q=\pi=2\kF$ is a result of the complete softening of the
renormalized phonon dispersion and is characteristic of the displacive
Peierls transition.\cite{HoWeBiAlFe06} 

\begin{figure}
 \includegraphics[width=0.45\textwidth]{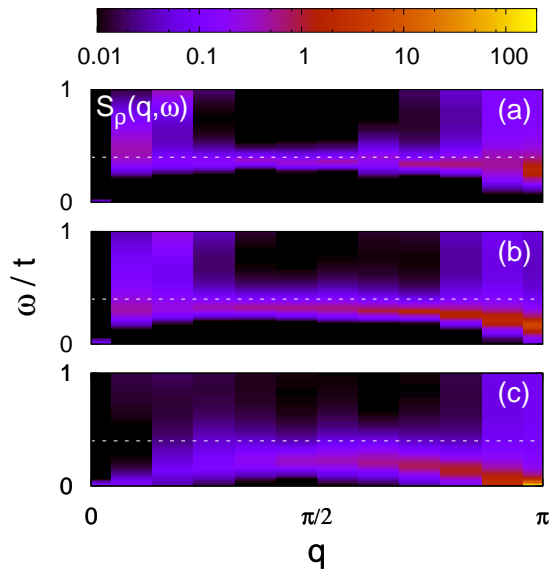}
  \caption{\label{fig:dynamics_softening} (Color online) Dynamic charge structure
    factor $S_\rho(q,\om)$ at low energies.
    Horizontal lines indicate the bare phonon frequency $\om_0$. 
    Here $\om_0=0.4t$, $\beta t=N=22$ and (a) $\lambda=0.25$ (LL), (b)
    $\lambda=0.5$ (LL), and (c) $\lambda=1$ (PI).}
\end{figure}

The renormalization of the phonon mode with increasing $\lambda$ is well
visible in Fig.~\ref{fig:dynamics_softening}, which shows the low-energy
range of $S_\rho(q,\om)$ on a larger scale.  Already for $\lambda=0.25$
[Fig.~\ref{fig:dynamics_softening}(a)] a partial softening is
visible close to $q=\pi$. This trend continues with increasing $\lambda$,
until the phonon has become gapless at $q=\pi$ in the Peierls state
[Fig.~\ref{fig:dynamics_softening}(c)]. The softening scenario agrees well
with previous work.\cite{HoWeBiAlFe06,CrSaCa05,SyHuBeWeFe04,SyHueBe06} From
$S_\rho(q,\om)$ we cannot detect the previously reported
splitting of the phonon mode into two branches, one of which hardens toward
$\om_0$ with increasing $\lambda$;\cite{HoWeBiAlFe06,CrSaCa05} this may be
a result of the very small spectral weight of the upper branch in the
adiabatic regime.

The linear long-wavelength mode characteristic of the metallic phase
[Fig.~\ref{fig:dynamics_LL}(a)] is completely suppressed with
increasing $\lambda$, see Fig.~\ref{fig:dynamics_softening}. From
Fig.~\ref{fig:dynamics_Peierls}(a), we can infer a gap of about $0.75t$
(left arrow) above which the particle-hole continuum starts. However, we can also
estimate a second characteristic energy of about $2t$ (right arrow) where the spectral
weight of the continuum increases noticeably. These two energy scales are
related to the single-particle spectrum below. Our results for
the dynamic charge structure factor are in accordance with previous
calculations of the phonon spectral function for the present
model.\cite{HoWeBiAlFe06,SyHueBe06} By comparison,
the Luther-Emery phase of the spinful Holstein model
is characterized by a soft phonon mode in addition
to the particle-hole continuum which remains gapless for $q\to0$.\cite{assaad:155124}

The single-particle spectral function in the Peierls state is
shown in  Fig.~\ref{fig:dynamics_Peierls}(b). As hinted at by the results
for $S_\rho(q,\om)$, it is found to consist of two
sets of features. There is a main, high-energy band that follows the free
dispersion far away from $\kF$, and is split by the {\em mean-field gap} (see below)
$\Delta_\text{mf}\approx \Ep=2t$ (indicated by the arrow). This value agrees with the
gap $2t$ discussed above. The main band also reveals backfolded shadow bands
for $q>\kF$ and $\om<\om_0$ respectively for $q<\kF$ and
$\om>\om_0$. Additionally, within the mean-field gap, we have dispersive
low-energy modes with a smaller gap $\Delta\approx0.75t$ at $\kF$. We refer
to this actual gap (determining the low-energy properties of the system) as the {\em Peierls gap}.

To first understand the high-energy features of $A(q,\om)$, we
consider the adiabatic limit $\om_0=0$ of the spinless Holstein model,\cite{HiFr83II}
\begin{equation}
\hat{H}' = \hat{H}_0 - \oh K\sum_i Q_i^2 - \gamma \sum_i Q_i (\on_i - \oh)\,.
\end{equation}
With the mean-field ansatz $Q_i=(-1)^i \xi$, we obtain the bands
$E_{\pm}(q)=\pm[\epsilon(q)^2 + \Delta_\text{mf}^2/4]^{1/2}$ with a gap
$\Delta_\text{mf}$ at $\kF=\pi/2$. The mean-field gap is related to the
lattice order parameter, $\Delta_\text{mf}\sim \xi^2$, and as for
displaced oscillator states the lattice shift on occupied sites is
$\xi\sim\sqrt{\Ep}$ so that we expect $\Delta_\text{mf}\sim\Ep$, in agreement
with the QMC results. Rather than solving the gap equation, we simply set
$\Delta_\text{mf}$ equal to the gap in our numerical data.

The resulting single-particle spectral
function\cite{Vo.Pe.Zw.Be.Ma.Gr.Ho.Gr.00} contains two
branches following $E_\pm(q)$, centered around $q=0$ respectively
$\pi$, with spectral weights $w_\pm(q)=[1\pm\epsilon(q)/|E_\pm(q)|]/2$, and is
shown in Fig.~\ref{fig:dynamics_Peierls_static}. Setting
$\Delta_\text{mf}=\Ep$ and replacing the $\delta$ functions by Lorentzians
with scaling parameter $0.3t$, we find excellent agreement with the
high-energy features in Fig.~\ref{fig:dynamics_Peierls}(b). In particular,
the backfolded shadow bands (including the $q$-dependence of the spectral
weight) emerge as a natural feature of gapped systems with competing periodic
potentials.\cite{Vo.Pe.Zw.Be.Ma.Gr.Ho.Gr.00} The fact that the high-energy
features of the quantum case $\om_0>0$ are well captured by mean-field theory
justifies the notation $\Delta_\text{mf}$ introduced above. 

\begin{figure}
 \includegraphics[width=0.45\textwidth,clip]{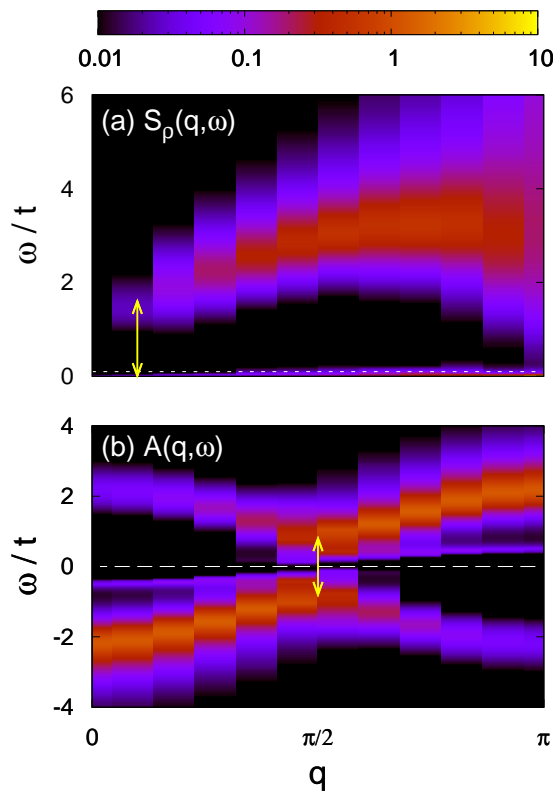}
  \caption{\label{fig:dynamics_Peierls_om0.1}     
  (Color online) (a) Dynamic charge structure
  factor $S_\rho(q,\om)$ and (b) single-particle spectral function $A(q,\om)$
  in the {\it Peierls phase}. Horizontal lines indicate (a) the
  bare phonon frequency $\om_0$, (b) the chemical potential $\mu=0$.
  Arrows designate the mean-field gap
  $\Delta_\text{mf}=\Ep=1.6t$. In (a), we have limited the intensity
  range to reveal the subdominant features.
  Parameters are  $\om_0=0.1t$, $\lambda=0.8$, and $\beta t=N=22$.}
\end{figure}

Let us now focus on the low-energy features inside the mean-field gap
[Fig.~\ref{fig:dynamics_Peierls}(c)]. The absence of the latter
in the mean-field results for the static limit suggests that they
are related to quantum phonon fluctuations. Indeed, a particle added 
to the half-filled Peierls state [as described by $A(q\geq\kF,\om\geq0)$,
cf. Eq.~(\ref{eq:akw})] is enabled by lattice fluctuations to propagate as a
polaron. For the present parameters, the dimerization of the lattice opens a
small but finite gap $\Delta\ll \Delta_\text{mf}$ in the corresponding polaronic band.
Support for this picture comes from several directions. First, the width of
the polaron bands in Fig.~\ref{fig:dynamics_Peierls}(c) is about 0.2t, in
excellent agreement with the single-polaron results for the same
parameters.\cite{HoHaWeFe06} It also shows the well known flattening
of the dispersion with increasing $q$.\cite{WeFe97,HoAivdL03} 
To demonstrate
this agreement, we include in Fig.~\ref{fig:dynamics_Peierls}(c) the band
dispersion $\en_\text{p}(q)$ of a single polaron for the same parameters.\cite{LoHoAlFe06}
To compare to half filling, we shift the dispersion by $\Delta/2$, and map
it from $[0,\pi]$ to $[\kF,\pi]$. This yields excellent agreement for all $q$.
Second, these
bands have large electronic spectral weight near $q=\kF$ (corresponding to
$q=0$ for a single polaron) but very small weight far away from $\kF$; the
character of the coherent quasi-particle changes from electronic to
phononic when the band energy intersects the bare phonon
frequency,\cite{WeFe97,HoAivdL03} see also Fig.~\ref{fig:dynamics_LL2}. Third, the energy gap $\Delta$ of the
polaron bands, which is a property of the many-particle Peierls state, matches
the corresponding excitation in the phonon spectral function for the same
parameters,\cite{HoWeBiAlFe06} and agrees well with the phonon signature in
Fig.~\ref{fig:dynamics_softening}(c) near $q=\pi/2$. For a single polaron,
no phonon renormalization occurs, and the polaron band dispersion is clearly
visible in the phonon spectral function.\cite{LoHoAlFe06} 
The spectral function for the same
parameters as Fig.~\ref{fig:dynamics_Peierls} has been calculated by cluster
perturbation theory.\cite{HoWeBiAlFe06} Whereas the polaron bands are
difficult to identify in these approximate results, their signature
inside the mean-field gap is clearly visible in the exact density of
states.\cite{WeBiHoScFe05} The QMC results are in excellent agreement with
the $T=0$ wavevector-resolved spectral function for these parameters obtained by exact
diagonalization.\cite{Wellein} An alternative  explanation for the polaron
bands in terms of thermal excitations based on QMC simulations at much higher
temperature was given before.\cite{HoNevdLWeLoFe04} Polaron bands inside the
static mean-field gap were also predicted analytically,\cite{Brazovskii78}
and may be related to soliton excitations.\cite{Heeger88}

From LL theory, the existence of the low-energy polaron modes is a
necessary consequence of the continuous nature of the Peierls transition. The
hybridized charge and phonon modes of the metallic phase evolve continuously
with the coupling strength $\lambda$. One of the resulting two modes
represents the backfolded high-energy band, whereas the other mode is
the polaron band. This structure is already visible in the LL phase, see
Figs.~\ref{fig:dynamics_LL}(b) and~\ref{fig:dynamics_LL2}. The Peierls
gap $\Delta$ in the polaron band opens exponentially slowly at the
KT transition.

In the light of previous work on many-polaron systems, the validity of the
single-polaron picture at half filling is not obvious. 
Away from half filling, the spinless Holstein model is metallic,
and the overlap of the (extended) lattice distortions of individual carriers
has been found to lead to a renormalization toward weak coupling and a loss
of polaronic signatures.\cite{HoNevdLWeLoFe04,LoHoFe06,HoHaWeFe06} However,
the electrons constituting the insulating Peierls state are ordered in
the dimerized lattice potential. Hence, an additional particle or hole 
does not undergo screening and behaves as a polaron, which moves in a
dimerized but fluctuating potential. The most obvious impact
of this Peierls background is the opening of the gap $\Delta$.

To complete the understanding of the dynamic charge correlations, we consider
the strong-coupling regime. From single-polaron theory, we expect a strong
reduction of the polaron bandwidth and the electronic spectral weight; the
latter also becomes almost independent of $q$.\cite{WeFe97,HoAivdL03} The
many-body Peierls gap $\Delta$ should increase with increasing dimerization
($\sim \sqrt{\Delta}$) of the lattice compared to phonon fluctuations
($\sim\sqrt{\om_0}$), leading to a suppression of the coherent polaron motion.

Exact $T=0$ results for the single-particle spectrum in the Peierls phase
have been reported by Sykora \etal,\cite{SyHuBeWeFe04} for $N=8$,
$\om_0=0.1t$ and $\lambda=0.8$. On the linear scale of their Fig.~4, no
features of the kind discussed here are visible. These results fall into
the strong-coupling regime with respect to the Peierls transition,
$\lambda\approx2\lambda_\text{c}$. In particular, the mean-field gap
$\Delta_\text{mf}\gg\om_0$. As demonstrated by the QMC results for the parameters of Sykora
\etal\cite{SyHuBeWeFe04} shown in Fig.~\ref{fig:dynamics_Peierls_om0.1}(b),
the spectral weight of the polaron band is extremely small. At $\kF$, the
gap $\Delta$ almost coincides with the mean-field gap $\Delta_\text{mf}$.
The spectrum is dominated by the mean-field features of
Fig.~\ref{fig:dynamics_Peierls_static}. As a result, the dynamic charge
structure factor in Fig.~\ref{fig:dynamics_Peierls_om0.1}(a) reveals only one
energy scale ($\Delta_\text{mf}\approx\Delta$) for the onset of the particle-hole
continuum, in addition to the signatures of phonon softening at low energies.
A
similar result is expected for $\om_0=0.4t$ at much stronger coupling where
the CTQMC method becomes inefficient. 

Finally, the features of the spectral function in the Peierls state discussed
here are specific to the adiabatic regime $\om_0< t$. In the nonadiabatic
case $\om_0\gg t$, the critical coupling for the Peierls transition is large,
so that electrons become heavy, small polarons already below $\lambda_\text{c}$.\cite{HoWeBiAlFe06}
The corresponding polaron bands are extremely narrow and have very small
electronic spectral weight.\cite{WeFe97,HoAivdL03} Similarly, the dispersion
of the high-energy features is also substantially reduced. 

\section{Conclusions}\label{sec:conclusions}

We have applied the exact continuous-time Monte Carlo method to
the spinless Holstein model in the adiabatic regime. We find that the method
is well suited to study the Peierls metal-insulator transition on the
lattice. 
We have complemented previous work by computing the dynamic charge
structure factor, and showing that is allows to track the softening of the
phonon mode and the appearance of a charge gap with increasing
electron-phonon coupling. 
For the single-particle spectral function, we explicitly demonstrated that
the mixed charge and phonon modes of Luttinger liquid theory constitute the
spectrum both below and above the Peierls transition. Most importantly, the
spectrum in the Peierls phase consists in general of high-energy features
readily understood in the static limit, and gapped low-energy polaron bands 
originating from quantum lattice fluctuations. These modes disappear in the
strong-coupling regime.
Our findings reconcile and unify previous and present
results with both polaron theory and Luttinger liquid theory.

The spinless Holstein model captures two essential consequences of
electron-phonon coupling, namely the formation of polarons (dressed
quasi-particles) and the Peierls metal-insulator transition at commensurate filling.
For any finite electron-phonon coupling, the electrons acquire some polaronic
character. Based on this and previous work, we can
conclude that starting from the low-density limit the well-defined polaron signatures known from the
single-electron case\cite{HoAivdL03} first become washed out with increasing
density in the metallic phase.\cite{HoHaWeFe06} However, once we approach the
Peierls phase at half filling, mutual screening is strongly suppressed due to
charge order and lattice dimerization, and
clear polaron signatures reappear in the single-particle spectrum.

\vspace*{-1em}
{\begin{acknowledgments}%
  We are grateful to Satoshi Ejima and Gerhard Wellein for valuable discussions.
  This work was supported by the DFG through FOR1162, and KONWHIR Bavaria.
  The generous computer time at the J\"ulich Supercomputing Centre is acknowledged.
\end{acknowledgments}}


\end{document}